\documentclass[10pt,a4paper]{article}
\usepackage{latexsym,epsfig,multicol}
\usepackage{color}
\usepackage{graphicx}
\usepackage{pifont}
\usepackage{verbatim}

\def\beq{\begin{eqnarray}}
\def\eq{\end{eqnarray}}

\def\beqn{\begin{eqnarray*}}
\def\eqn{\end{eqnarray*}}



\usepackage{amssymb}
\usepackage{amsfonts}
\usepackage{amsmath}%
\usepackage{graphicx}

\providecommand{\U}[1]{\protect\rule{.1in}{.1in}}

\title{ Right-handed heavy neutrinos in the littlest Higgs model}
\author{F.~M.~L.~de Almeida Jr., Y.~A.~Coutinho, \\
J.~A.~Martins Sim\~oes, A.~J.~Ramalho, S.~Wulck\\
Instituto de F\'isica, \\
Universidade Federal do Rio de Janeiro,\\
Rio de Janeiro, RJ, Brazil\\
\\
M.~A.~B.~do Vale\\
Departamento de Ci\^encias Naturais,\\
Universidade Federal de S\~ao Jo\~ao del Rei, \\
S\~ao Jo\~ao del Rei, MG, Brazil }

\date{}

\begin{document}

\maketitle

\begin{abstract}
In this paper we discuss the consequences of including  new heavy right-handed neutrino singlets $N_R$ in the littlest Higgs model. These new states are not connected with the light neutrinos {\it via} the seesaw mechanism. A very interesting property of this extended model is the full coupling of the new neutral gauge boson $A_H$ to the right-handed neutrinos $N_R$, giving large total cross sections and suggesting a wide range of experimental search for the $N_R$ at the CERN-LHC pp collider and future ILC electron-positron collider.

\vskip 2cm 
PACS: 12.60.-i, 14.60.Pq, 14.60.St
\vskip 1cm
\par
\noindent
email:marroqui@if.ufrj.br, yara@if.ufrj.br, simoes@if.ufrj.br, ramalho@if.ufrj.br, steniow@if.ufrj.br,aline@ufsj.edu.br

\end{abstract}
\eject

\par
Present day particle physics faces at least two  problems with the masses of some of the fundamental particles.
The first one is that the high-energy loop corrections to the Standard Model Higgs mass  suggest that new physics must be added to stabilize quantum corrections. This is the main motivation for supersymmetry and for the "little Higgs" scenario
 \cite{AHN}. The other problem is the experimental observation of  very small neutrino masses \cite{PDG}. The pattern of  small neutrino masses and high mixing angles still waits for a deeper theoretical understanding.
\par
In this paper we point out a possible way to circumvent these problems: the inclusion of right-handed singlet neutrinos in the littlest Higgs model (LHM). There is a recent proposal in this direction \cite{ASM}, but using the enlarged little Higgs group $SU(3)_W \otimes U(1)_Y$. Another proposal was put forward in \cite{THB}. These authors considered the seesaw mechanism for generating neutrino masses. This mechanism implies new mass scales well above the little Higgs cutoff $\Lambda \simeq 4\pi f \simeq 10$ TeV and their conclusions were that this approach should be avoided. Since the seesaw mechanism has not been experimentally confirmed yet \cite{BNS}, it must be considered as one possible scenario, but not the only one. In fact, in the last few years a number of alternative models for neutrino masses have been proposed \cite{ALM,CHA,CEC,BUC,GON}.
These models can differ in the fundamental mechanism for giving a small neutrino mass. They can include either more isosinglet right-handed neutrino states or unified theories with $B-L$ breaking or some singular mass matrix and  new symmetries. But they all have one fundamental point in common: they do not need an ultra-high scale as in the seesaw mechanism. This property makes all of them suitable candidates for the fermionic sector of LHM, since they do not imply new scales above the little Higgs cutoff $\Lambda \simeq 10$ TeV, as already mentioned. The inclusion of new isosinglet right-handed neutrino states follows the general pattern of the other charged fermions of the Standard Model (SM) and, if they include Majorana mass terms, they offer a natural explanation for a possible  lepton number violation.
\par
As an example of such models we consider a scenario with three left-handed leptonic doublets and two new right-handed neutrino singlets. The mass Lagrangian for this model is given by
\par
\begin{equation}
{\cal L}_{mass} =  \sum^{3}_{i=1} \sum^{2}_{j=1} a_{ij} \left(\bar l_i ,\, \bar{\nu_l}_{i}\right)_L H N_{R_{j}}+ \frac{1}{2}  \sum^{2}_{i=1} \sum^{2}_{j=1}  M_{ij} \bar N^c_{R_{i}} N_{R_{j}}
\end{equation} 
where we have included Majorana mass terms for the right-handed singlets and the left-handed leptonic doublets are coupled to the standard model Higgs field $H$. If we consider all the couplings of the left-handed leptons to be of the order of the Fermi scale $v_F$ and allow a small mixing in the right-handed sector, then we have the following mass matrix,
\beq
{\cal M} =\left(\begin{array}{llllll}
0 & 0 & 0 &  {v}_F &  {v}_F \\
0 & 0 & 0 &  {v}_F & {v}_F\\
0 & 0 & 0 &  {v}_F &{v}_F \\
 {v}_F &  {v}_F &  {v}_F & M & M ^{\prime} \\
 {v}_F &  {v}_F & {v}_F & M ^{\prime}  & M \\
\end{array}\right).
\eq

\par
 The diagonalization of this matrix leads to two states with zero mass, one with 
$\Delta M = M - M^{\prime}$ and two heavy states with masses given by:
\begin{equation}
M_{1,2} =  \frac{\sqrt{( M+M^{\prime})^2+24 v_F^2}\pm  (M+M^{\prime})}{2} 
\end{equation} 

\par
Two small neutrino masses could be radiatively generated and a third one is a consequence of a small $ \Delta M= M-M^{\prime}$. If we now make the additional hypothesis that the Majorana scale is $ M \simeq M' \simeq 1$ TeV then the heavy mass states are $ M_1 \simeq 170$ GeV and $M_2 \simeq  2200$ GeV.
\par
The important point in this model is that if the neutrino mass mechanism generation is not of the seesaw type, then we have no physical mass scales above the little Higgs cut-off at $10$ GeV. The new possible states can be experimentally searched in future p p  and $e^+$ $e^-$ colliders. Our example is not the only model possible and there are many other models with this property as given in references \cite{ALM,CHA,CEC,BUC,GON}.
\par
We turn now our attention to the inclusion of these new right-handed singlets in the littlest Higgs model.
The littlest Higgs gauge interactions come from the breaking of $SU(5)$ into its subgroup $[SU(2) \otimes  U(1)]^2$.
The two stages of the $[SU(2) \times U(1)]^2$ breaking give us first the SM $SU(2)_L \otimes U(1)_Y$ group at the scale $f$ of a few  TeV and then $U(1)_{\text em}$ at the Fermi scale $v_F \simeq 246$ GeV.

The Lagrangian of the gauged theory comes from the derivative replacement:
\begin{equation}
\partial_{\mu} \Sigma \longrightarrow D_{\mu}\Sigma =\partial_{\mu}\Sigma - i \sum_{j=1} ^{2} \left[ g_j W^a_{j\mu}\left(Q^a_j\Sigma + \Sigma Q^{aT}_j\right) +
g^{\prime}_j B_{j\mu}\left( Y_j \Sigma+\Sigma Y_j \right) \right], 
\end{equation} 
\noindent
where $g_j$ and  $g^{\prime}_j$ are the coupling constants associated to the gauge fields $W^a_j$ and $B_j$, $Q^a$ are the group generators.   

\par
The linear combinations of gauge fields that acquire mass in the TeV scale are

\begin{eqnarray}
W^{\prime \mu} = -\cos \eta \ W_1^{\mu} + \sin \eta \ W_2^{\mu}\nonumber \\
B^{\prime \mu} = - \cos \eta^{\prime} B_1^{\mu} + \sin \eta^{\prime}B_2^{\mu} 
\end{eqnarray} 

\noindent whereas the combinations

\begin{eqnarray}
W^{\mu} = \sin\eta \ W_1^{\mu} + \cos\eta \ W_2^{\mu} \nonumber \\
B^{\mu}= \sin\eta^{\prime} B_1^{\mu}+ \cos\eta^{\prime}B_2^{\mu}
\end{eqnarray} 
\noindent remain massless.

\par
The mixing angles are given by

\begin{eqnarray}
s \equiv \sin\eta = \frac{g_2}{\sqrt{g_1^2+g_2^2}} & {\text and }&
\  s^{\prime}\equiv \sin^{\prime}\eta = \frac{g_2^{\prime}}{\sqrt{g_1^{\prime \ 2}+g_2^{\prime \ 2}}}.
\end{eqnarray} 

\par
The massless (at this stage) SM gauge bosons $W$ and $B$ have couplings

\begin{eqnarray}
g = g_1 \sin\eta= g_2 \cos\eta & and & \ g^{\prime}= g_1^{\prime} \sin\eta^{\prime} = g_2^{\prime} \cos\eta^{\prime}. 
\end{eqnarray}

\par
The inclusion of right-handed neutrino states can be easily achieved, but with the following essential remark: their $U(1)$ hypercharges $Y_1$ and $Y_2$ do not need both to be zero, but only their sum

\begin{eqnarray}
Y = Y_1 + Y_2 \equiv 0
\end{eqnarray} 
must be zero.
\par
Then it follows that the new isosinglet gauge interactions in the neutral gauge boson sector are given by

\begin{equation}
{\cal L}_{gauge}\left(N_R\right) =\bar N_R^0 \gamma^{\mu}\left(\partial_{\mu}- 
i g_1^{\prime} B_{1\mu}Y_1-ig_2^{\prime}B_{2\mu}Y_2\right) N_R^0,
\end{equation} 
with the usual SM definition for the electroweak mixing angle,

\begin{equation}
s_W \equiv \sin\theta_W  = \frac{g^{\prime}} {\sqrt{g^2+g^{\prime \ 2}}}
\end{equation} 
\noindent and $c_W \equiv \cos \theta_W$. Neglecting $v^2/f^2$ terms we have the LHM neutral gauge boson sector:

\begin{eqnarray}
&&A_L^{\mu} = s_W W^{3 \mu} + c_W B^{\mu} ,\\ \nonumber
&&Z_L^{\mu} = c_W W^{ 3\mu} - s_W B^{\mu} ,\\ \nonumber
&&A_H^{\mu}= B^{\prime \mu},\\ \nonumber
&&Z_H^{\mu} = W_3^{\prime \mu},
\end{eqnarray}

\noindent where $A_L$ and $Z_L$ are the SM neutral gauge bosons (photon and $Z^0$), $A_H$ and $Z_H$ two heavy neutral gauge bosons. $A_H$ is new lightest boson in the LHM and it is expected to be detected at the LHC energies, if it exists.

\par
The Lagrangian related to the neutral interactions for $N^0_R$ has the following combination:
\begin{equation}
g_1^{\prime} B_1^{\mu} Y_1 + g_2^{\prime} B_2^{\mu} Y_2 = g_1^{\prime} s^{\prime}\left(Y_1+Y_2 \right)\left( c_W A^{\mu} - s_W Z_L^{\mu} \right) + A_H^{\mu} \frac{g^{\prime}}{s^{\prime}c^{\prime}}Y_1\left( c^{\prime 2}- s^{\prime 2}\right). 
\end{equation}

Clearly the condition $Y_1+Y_2 = 0$ not only decouples the $N_R$ state from the photon but also from the $Z_L$. This is a consequence of the standard breaking of the LHM. This aspect eliminates the
strong constraint from the $Z_L$ width that restricts the right-handed neutrino mass in other models \cite{HUG}.

\par
Before writing the final interactions for $N^0_R$, we must proceed to the neutrino mixing through the rotation matrix,

\begin{eqnarray}
\left(\begin{array}{l}
N_0 \\
\nu_0
\end{array}\right) & = &\left(\begin{array}{ll}
\cos\theta_{\text {mix}} & -\sin\theta_{\text {mix}}  \\ 
\sin\theta_{\text {mix}} & \phantom{-} \cos\theta_{\text{mix}} \end{array}\right)
\left(\begin{array}{l}
N \\
\nu_e
\end{array}\right).
\end{eqnarray}

\par
The agreement of experimental data and the SM predictions implies $sin^2 \theta_{mix}$ $\lesssim 10^{-3}-10^{-4}$ \cite{ALM}. This small mixing can suppress heavy neutrino pair production in many models, as noticed some time ago \cite{ALM}. But this is not the case for the LHM with isosinglets $N_R$, from now on called LHM$_{DS}$. The present experimental bound on new heavy neutral lepton mass is $M_N > 100$ GeV \cite{PDG}.
\par
The final neutral interaction involving $N_R$ and considering neutrino mixing is described by
\begin{equation}
{\cal L}_{gauge}\left( N_R\right) =\frac{g^{\prime}}{2s^{\prime} c^{\prime}}
\left( c^{\prime \ 2}- s^{\prime \ 2}\right)Y_1 {A_H}_{\mu} \bar N_R \gamma^{\mu}N_R \cos^2\theta_{mix},
\end{equation} 

\noindent where we can see explicitly the full coupling in the vertex $A_H\bar N_R N_R$.

\par
This is not the complete Lagrangian. There are other terms coming from the $(\nu_e \  e)_L$ doublet \cite{WAN}, but the $Z_L \bar N_L N_L $ vertex is suppressed by $sin^2 \theta_{mix}$. In order to study the consequences of introducing a new right-handed neutrino singlet, we present in the following figures the estimates for pair production of heavy neutrinos  in the LHM that comes only from the $(\nu_e \ e)_L$ doublet model and the case with new right-handed singlets. Since we have lepton number violation in this model, we can rotate the Dirac spinors of the right-handed neutrino singlets in two Majorana spinors \cite{CLI}, giving lepton number violation and the possibility of leptogenesis \cite{NIR} in the LHM. The other lepton and quark couplings are the same in both models.

\par
Now we turn our attention to the total cross sections for the production of a pair of heavy Majorana neutrinos in the reactions $e^+ + e^- \longrightarrow N + N$ at the ILC and $p + p \longrightarrow N + N + X$ at the LHC. 
We have compared the results from the traditional LHM, that we call LHM$_{D}$ which has a heavy  neutrino $N_L$, with the LHM$_{DS}$ which has a new heavy right-handed neutrino singlet $N_R$.
Both models have been implemented in CompHEP \cite{COM} and the total cross sections have been calculated using
CTEQ5L parton distribution functions \cite{CTQ}. The heavy Majorana neutrino can decay to $W^+ + \ell^-$, $W^- + \ell^+$,
$Z_L + \nu_{\ell}$, where $\ell = e, \mu,\tau$, with branching ratios dependent on $M_N$ \cite{AGU}.
The main final states originating from the production of a pair of Majorana neutrinos will be $\ell^\pm + \ell^{\prime \pm} + W^\mp + W^\mp$, where $\ell ^{\prime}= e, \mu,\tau$, like-sign dilepton events plus jets and leptons, with or without missing energy. In our calculations, we have considered $sin^2 \theta_{\text{R mix}} = 0.0052$ \cite{FML}, $c^ \prime$ in the range $0.62 - 0.73$, allowed by the electroweak precision data \cite{CSA, GRE} and $M_{A_H}$ and $M_N$ as free parameters.

\par
The contributions to the process $e^+ + e^- \longrightarrow N + N$ come from $Z_L$, $Z_H$, $A_H$, $W^+$ and $V^+$ exchanges. The dominant diagram is given by the $A_H$ s-channel exchange.  
The total cross section enhancement is a consequence of the $cos^2 \theta_{mix}$ factor in the coupling of the $N_R$ to $A_H$. Confirming this expectation the results for a future $e^+e^-$ collider for energies $\sqrt s = 500$ GeV and $\sqrt s = 1$ TeV are shown in Figures 1 and 2 respectively for $M_N= 100 $ GeV and two values of $c^{\prime}=0.62$ and $0.73$. The total cross section difference between the two models is $\simeq 10^4$ in the resonant region. For $\sqrt s = 1$ TeV, these values are still of order $\simeq 10^3$ in central region too.
The total cross section enhancement due to the new heavy right-handed neutrino singlet would make this new neutrino observable up to $\sqrt s =M_{A_H}/2$, should this new neutral gauge boson really exist. The total cross sections for other values of $c^\prime$ are between the two extremes shown, except for $c^\prime = s^\prime$, where $A_H$ decouples from $N_R$, as can be seen from Eq. 13. It should be noted that this is not the same angle where $A_H$ decouples from the SM fermions \cite{WAN}. This point will be explored in more detail in a work in progress. Considering an annual luminosity ${\cal L}= 100 $ fb$^{-1}$, ILC can produce $10^8$ and $10^5$ events for $\sqrt s= 500$ GeV and $\sqrt s= 1$ TeV respectively, in the resonant region without cuts.
\par

The results for the proton-proton collisions at $\sqrt{s} = 14$ TeV, that will take place at the LHC, are shown in Figures 3 and 4 for two $M_{A_H}$ values. We obtained the total cross sections for neutrino pair production in the process $p + p \longrightarrow N + N + X$ as a function of the neutrino mass, taking a cut $M_{ij} > 100$ GeV on the invariant mass of final fermions, and for $c^ \prime$ in the range $0.62 - 0.73$. Considering an annual luminosity of ${\cal L}= 50 $ fb$^{-1}$ and $M_{A_H}= 500$ GeV, LHC can produce $500$ events for $M_N= 100$ GeV and $5$ events for $M_N \simeq 220$ GeV. 
For an $e^+e^-$ collider there is a large number of events at the resonance. At the LHC collider there is a wide range of neutrino mass that can be explored up to $M_N \simeq M_{A_H}/2$.

\begin{figure} 
\begin{center}
\includegraphics[width=.6\textwidth]{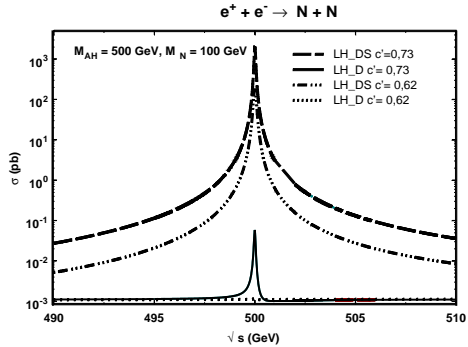}
\caption{Total cross section for $e^+ + e^- \longrightarrow N + N$ at the ILC for the LHM$_D$ and LHM$_{DS}$ as a function of $\sqrt{s}$, for $M_{A_H}= 500$ GeV, $M_N= 100$ GeV, and two values $c^{\prime}= 0.62$ and $c^{\prime}= 0.73$.}
\end{center} 
\end{figure}

\begin{figure} 
\begin{center}
\includegraphics[width=.6\textwidth]{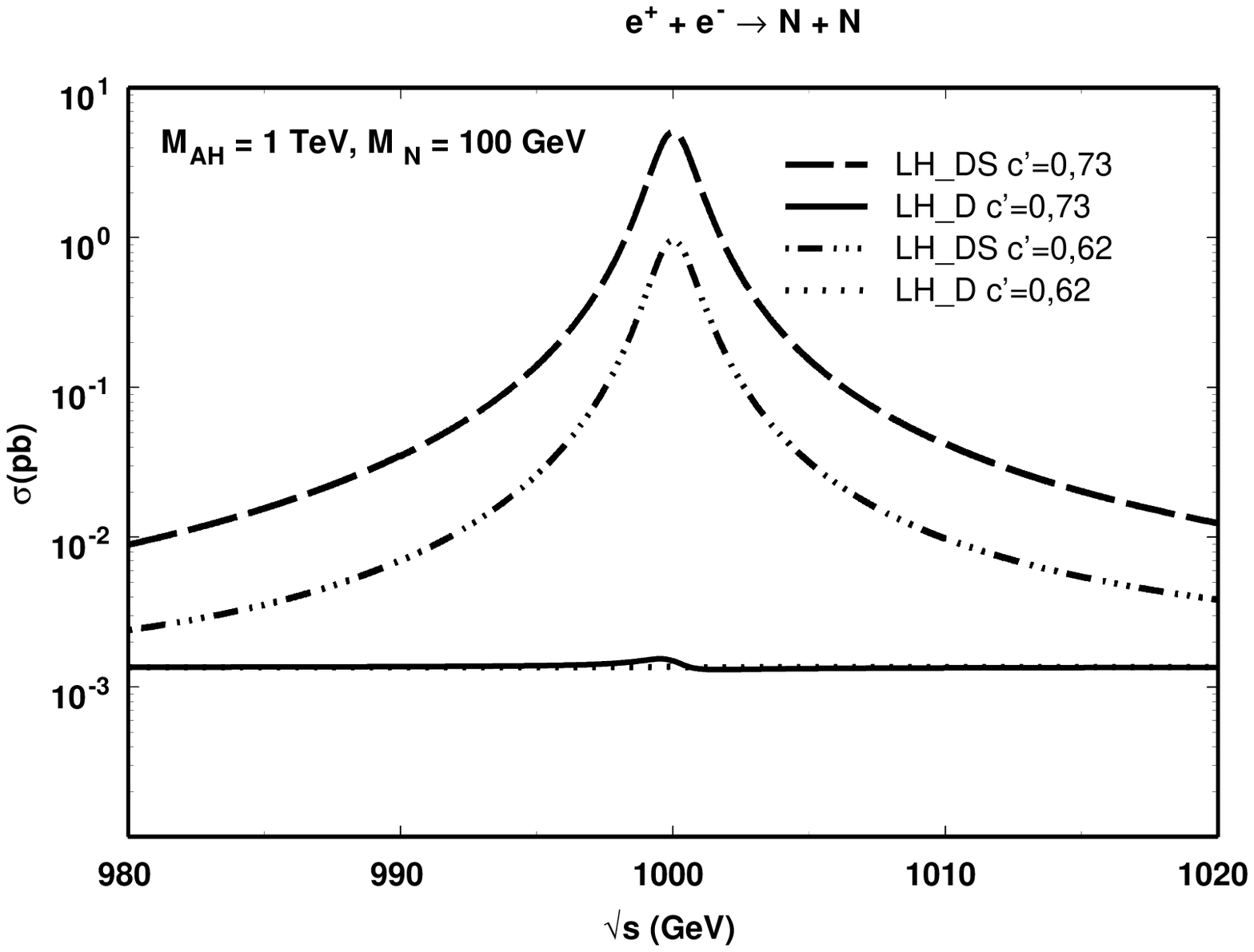}
\caption{Total cross section for $e^+ + e^- \longrightarrow N + N$ at the ILC for the LHM$_D$ and LHM$_{DS}$ as a function of $\sqrt{s}$, for $M_{A_H}= 1$ TeV,  $M_N= 100$ GeV, and two values $c^{\prime}= 0.62$ and $c^{\prime}= 0.73$.}
\end{center} 
\end{figure}

\begin{figure} 
\begin{center}
\includegraphics[width=.6\textwidth]{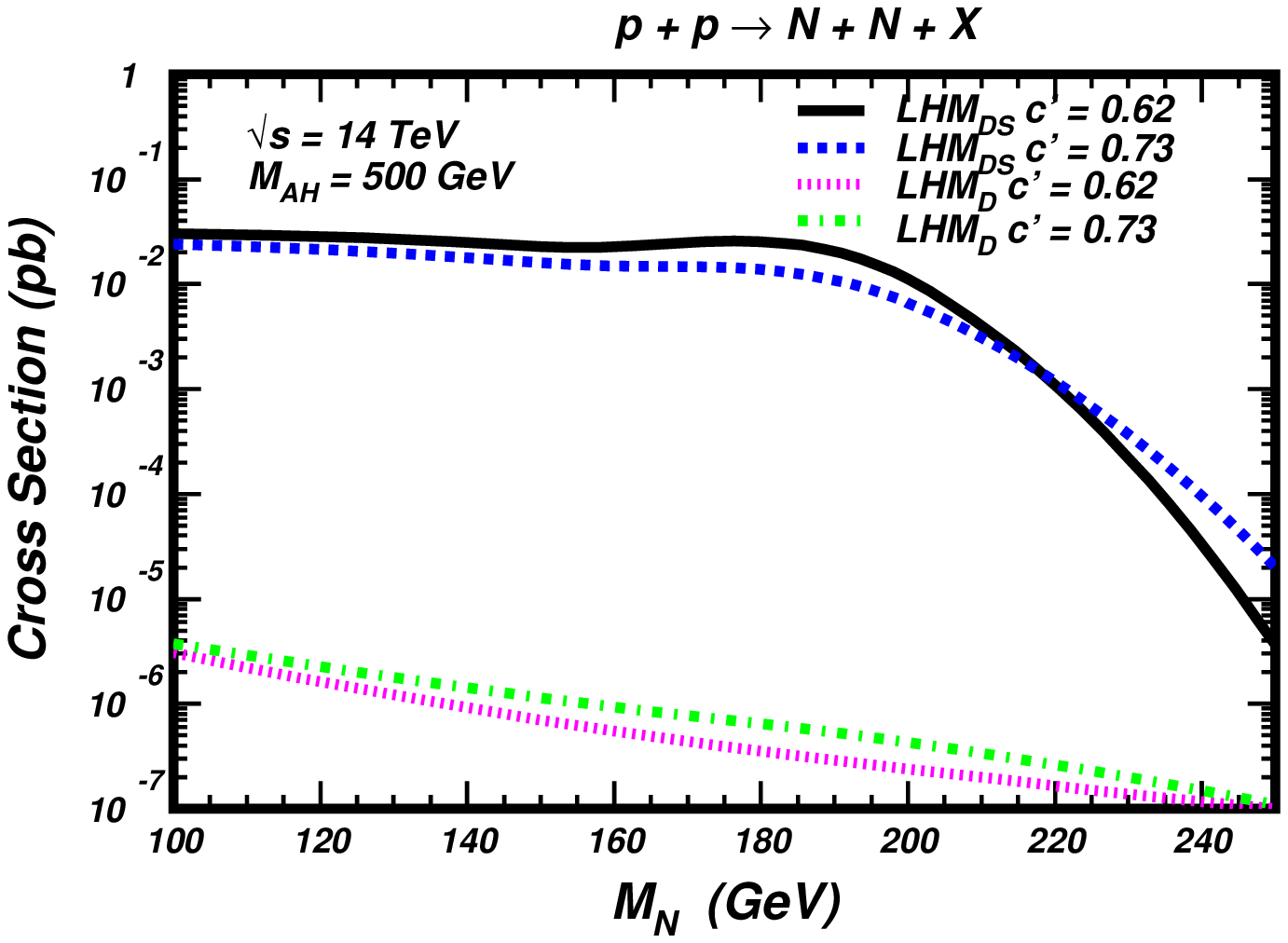}
\caption{Total cross section for $p + p \longrightarrow N + N + X$ at the LHC for the LHM$_D$ and LHM$_{DS}$ as a function of $M_N$, for $M_{A_H}= 500$ GeV, and two values $c^{\prime}= 0.62$ and $c^{\prime}= 0.73$.}
\end{center} 
\end{figure}

\begin{figure} 
\begin{center}
\includegraphics[width=.6\textwidth]{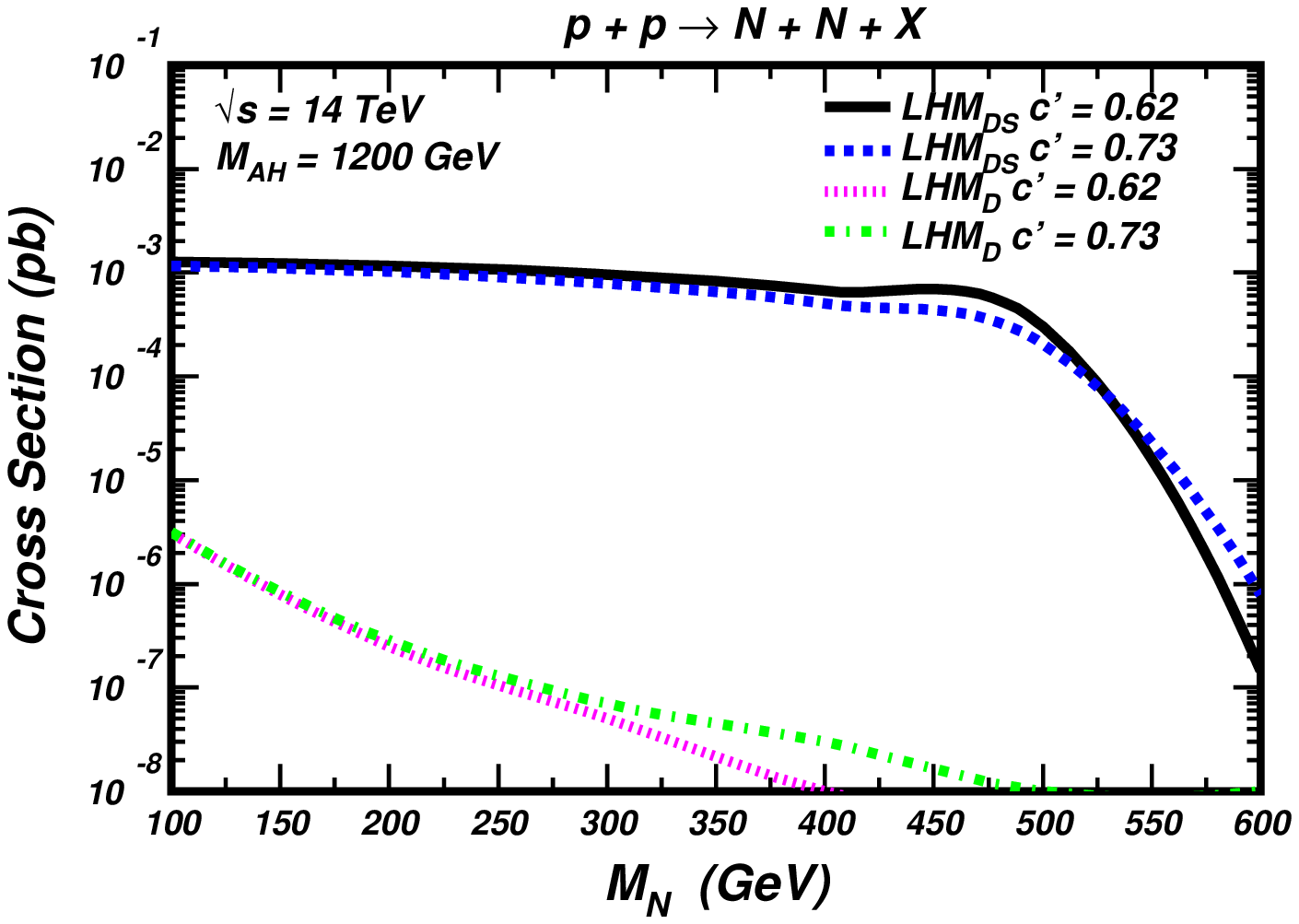}
\caption{Total cross section for $p + p \longrightarrow N + N + X$ at the LHC for the LHM$_D$ and LHM$_{DS}$ as a function of $M_N$, for $M_{A_H}= 1200$ GeV, and two values $c^{\prime}= 0.62$ and $c^{\prime}= 0.73$.}
\end{center}
\end{figure}

\par
We have shown that the introduction of  right-handed neutrino singlets $N_R$ in the content of the LHM presents a very particular behavior, leading to sizeable total cross sections in $e^+ + e^-$ and  p p collisions at the next generation of colliders and widening the $N_R$ mass search up to $M_{A_H}/2$. New charged heavy leptons in Little Higgs models are also possible \cite{YUE}. The associated production of
a new neutrino state $N_R$ $\it via$ $A_H$ exchange can be a very clear experimental test for the littlest Higgs model. An eventual lepton number violation would point to the Majorana nature of $N_R$ and this will be discussed in a separate publication.
\par
\vskip 1cm
\textit{Acknowledgments:} 
We thanks CNPq, FAPERJ and FAPEMIG for support.

\end{document}